# A modified porous titanium sheet prepared by plasma activated sintering for biomedical applications

Yukimichi Tamaki*, Won Sik Lee, Yu Kataoka, and Takashi Miyazaki

*Department of Oral Biomaterials and Technology, School of Dentistry, Showa University, 1-5-8 Hatanodai, Shinagawa-ku, Tokyo 142-8555, Japan*

**This study aimed to develop a contamination free porous titanium scaffold by a plasma activated sintering within an originally developed TiN coated graphite mold. The surface of porous titanium sheet with or without a coated graphite mold was characterized. The cell adhesion property of porous titanium sheet was also evaluated in this study. The peak of TiC was detected on the titanium sheet processed with the graphite mold without a TiN coating. Since the titanium fiber elements were directly in contact with the carbon graphite mold during processing, surface contamination was unavoidable event in this condition. The TiC peak was not detectable on the titanium sheet processed within the TiN coated carbon graphite mold. This modified plasma activated sintering with the TiN coated graphite mold would be useful to fabricate a contamination free titanium sheet. The number of adherent cells on the modified titanium sheet was greater than that of the bare titanium plate. Stress fiber formation and the extension of the cells were observed on the titanium sheets. This modified titanium sheet is expected to be a new tissue engineering material in orthopedic bone repair.**
* contact to: tamaki@dent.showa-u.ac.jp

## Introduction

A long segmental bone defects repair is one of the challenging problems in orthopaedic surgery. Although allogenic bone grafts are a current major option[1-3], this technique is associated with problems of significant failure rates, poor mechanical properties and immunological rejection[2]. Porous materials are of significant importance for bone tissue engineering applications because of the good biological fixation to surrounding tissue through bone tissue[4].

Porous titanium and titanium alloys have been investigated as they provide favorable mechanical properties with an elastic modulus closed to that of natural bone under a load bearing condition[5]. Surface characteristics of porous titanium are important determinants in its scaffold properties since the surface condition of titanium has been reported to play a critical role in bone formation associated with superior osteoblast adhesion and subsequent cell behaviors[6-10].

Recent studies have raised a concern on titanium-based biomateirlas that degradation of biological ability with increase in adsorption of organic impurities such as polycarbonyls and hydrocarbon onto the surface[6, 7]. This reduces hydrophilicity, adsorption of cell-binding proteins, and subsequent cell functions. Titanium-based biomaterials therefore need to be fabricated as contamination free surface.

There are a number of approaches in fabrication of porous titanium and titanium alloys such as sintering loose titanium powder or fibers, slurry sintering and also rapid prototyping[3, 4, 11, 12]. Initial surface contamination of sintered porous titanium would be unavoidable event as this is generally processed within the carbon graphite molds under a high thermal pressure[3, 4].

This study aimed to develop a contamination free porous titanium scaffold by a plasma activated sintering within an originally developed TiN coated graphite molds. The surfaces of porous titanium sheets with or without a coated graphite mold were characterized. The cell adhesion property of the porous titanium sheet was also evaluated in this study.

## Materials and Methods

*Specimen preparation*
JIS grade II titanium (KS-50, Kobe Steel, Tokyo, Japan) block was used as starting material. Narrow titanium fibers were made by turned with diameter approx. 0.4-0.5 mm. The surface of graphite molds were coated with thickness of 1.0 mm TiN by pre-sintering under vacuum.
Elemental titanium fibers were filled in the $\phi$ 30 mm x 1.0 mm graphite mold with or without a TiN coating. The titanium fibers were subjected to the pre-displacement with 1 MPa pressure in the graphite molds. The porous titanium sheet $\phi$ 30 mm x 0.5 mm was processed by plasma activated sintering under a vacuum at pressure of 20 MPa, at 3800 A for 15 sec.

*Scanning electron microscopy*
The surface topographies of the coatings on the specimens were then observed by SEM (S-2360N, Hitachi, Tokyo, Japan).

*Thin-film X-ray diffraction (TF-XRD)*
The crystalline phases of the titanium samples before the tests were detected by TF-XRD (XRD-6100, Shimadzu, Kyoto, Japan) with CuKα radiation. The XRD was operated at 40 kV and 40 mA with a scanning speed of 0.02°/4 s and a scanning range of 20–60°.

*Number of cells on titanium samples*
10 x 10 x 0.5 mm polished titanium plates and titanium sheets were subjected to cell culture. An osteoblastic cell line, MC3T3-E1 was obtained from the RIKEN Cell Bank (Tsukuba, Japan). Cells were cultured in α minimal essential medium (Gibco) containing 10% fetal bovine serum (Gibco) and 1% antibiotic (penicillin, Gibco) under a 5% $CO_2$ atmosphere at 37°C. Cells were suspended in the medium at $1 \times 10^5$ cells/mL and used for experiments. A 1-mL quantity of floating cells was cultured onto titanium samples at 37°C under 5% $CO_2$ for 1 d. A cell-counting kit (Dojindo, Kumamoto, Japan) was used for the measurement of cell adhesion. After incubation, each specimen was moved to another well and washed 3 times with PBS (Gibco) to remove non-adherent cells. Adherent cells were mixed with 1 mL of medium and 100 μL of reagent solution. After 1 hr of incubation, the absorbance at 450 nm was measured. The number of adherent cells was calculated from the activity of the original cell suspension.

*Stress Fiber Formation and Cell Morphology*
Specimens were placed in 24-well culture plates with 1 mL floating cells each. Subsequently, the specimens were incubated at 37°C in 5% $CO_2$ for 1 hr. Adherent cells on each specimen after 1 hr of cultivation were dehydrated after being washed with PBS. The cells were fixed with 3.7% formaldehyde in PBS and permeabilized by treatment with 0.1% Triton X-100 (Sigma, Tokyo, Japan) in PBS for 1 min. The cells were then incubated for 3 hrs in a rhodamine-conjugated phalloidin solution. After the cells were washed with water, stress fiber formation and cell morphology were observed with the use of a fluorescence microscope (E-600, Nikon, Tokyo, Japan).

*Statistical Analysis*
Results are expressed as mean ± SD (n=6) within each sample. The normal distribution of each value was confirmed using the Kolmogorov–Smirnov test. The appropriateness of the hypothesis of homogeneous variances was investigated by means of the Bartlett's test. Data were statistically analysed by ANOVA followed by a port-hoc Tukey test. A *p* value of less than 0.01 was considered significant.

**Results**

*Porous titanium sheet*
As shown in Fig.1, the porous titanium sheet was structured under the high thermal pressure.

*Surface characterization*
The titanium sheet prepared with the graphite mold showed distinctive TiC peaks while the sheet with coated graphite mold showed no carbon related impurities. The peaks of pure $TiO_2$ were only detected on the titanium sheet prepared with the coated graphite mold (Fig. 2).

*Adherent cells on titanium samples*
The number of adherent cells on titanium sheet was significantly ($p < 0.01$) higher than that of titanium plate after 1 d (Fig 3). Adherent cells on the titanium sheet had begun to showed stress fibers and widely extended, while the stress fiber formation and cell extension on titanium plate were not distinctive (Fig 4).

**Discussion**
The plasma activated sintering is a rapid sintering method associated with self-heating phenomena within the powder. This is capable of sintering metal or ceramic powders rapidly to its full density at a relatively lower temperature compared to the conventional furnace sintering methods. The carbon graphite mold has been employed in the plasma activated sintering due to its electro conductive property and thermostability[3, 4, 11, 13]. The direct heating of graphite mold and the large spark pulse current provide a very high thermal efficiency[14]. The peak of TiC was detected on the titanium sheet processed with graphite mold without TiN coating. Since the titanium fiber elements were directly in contact with the carbon graphite mold during processing, surface contamination is unavoidable event in this condition.
Alternatively, the TiC peak was not detectable on the titanium sheet processed within the TiN coated carbon graphite mold. This modified plasma activated sintering with the TiN coated graphite mold would be useful to fabricate the contamination free titanium sheet.
Recent study suggested that adsorption of organic impurities such as polycarbonyls and hydrocarbons on titanium surface are responsible for reducing the initial cells adhesion, subsequent proliferation and differentiation[6-8]. Amount of carbon absorbed on the titanium sheet seems to be an important part in determining the initial affinity level for osteoblasts and new bone formation.
The present study demonstrated that number of adherent cells on the modified titanium sheet was much grater than that of bare titanium plate.

Additionally, stress fiber formation and the extension of the cells were observed on the titanium sheets. The initial adhesion of cells induces stress fiber formation, phosphorylation of focal adhesion kinase and activation of other intracellular signal transduction molecules thereby affect cell proliferation, differentiation and new bone formation[15].Thus, the contamination free surface of modified titanium would be useful for new bone generation at a segmental bone defect in comparison with the unmodified sintered porous titanium.

In conclusion, the TiN coated carbon graphite mold is a new method for processing a contamination free porous titanium sheet. This modified titanium sheet is expected to be a new tissue engineering material in orthopedic bone repair.

**ACKNOWLEDGEMENTS**
This work was supported by MEXT, Haiteku (2009), a Grant-in-Aid for Scientific Research (B) from the Japan Society for the Promotion of Science, and a Grant-in-Aid for the Encouragement of Young Scientists (B) from The Ministry of Education, Culture, Sports, Science and Technology of Japan.

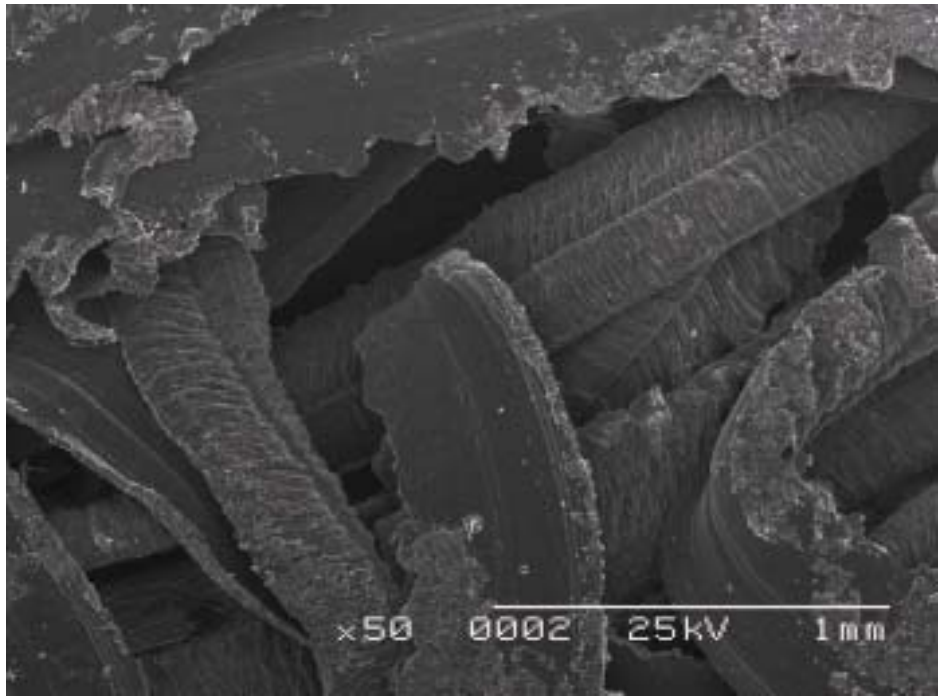

Figure 1 A representative SEM picture of the titanium sheet processed with a TiN coated graphite mold.

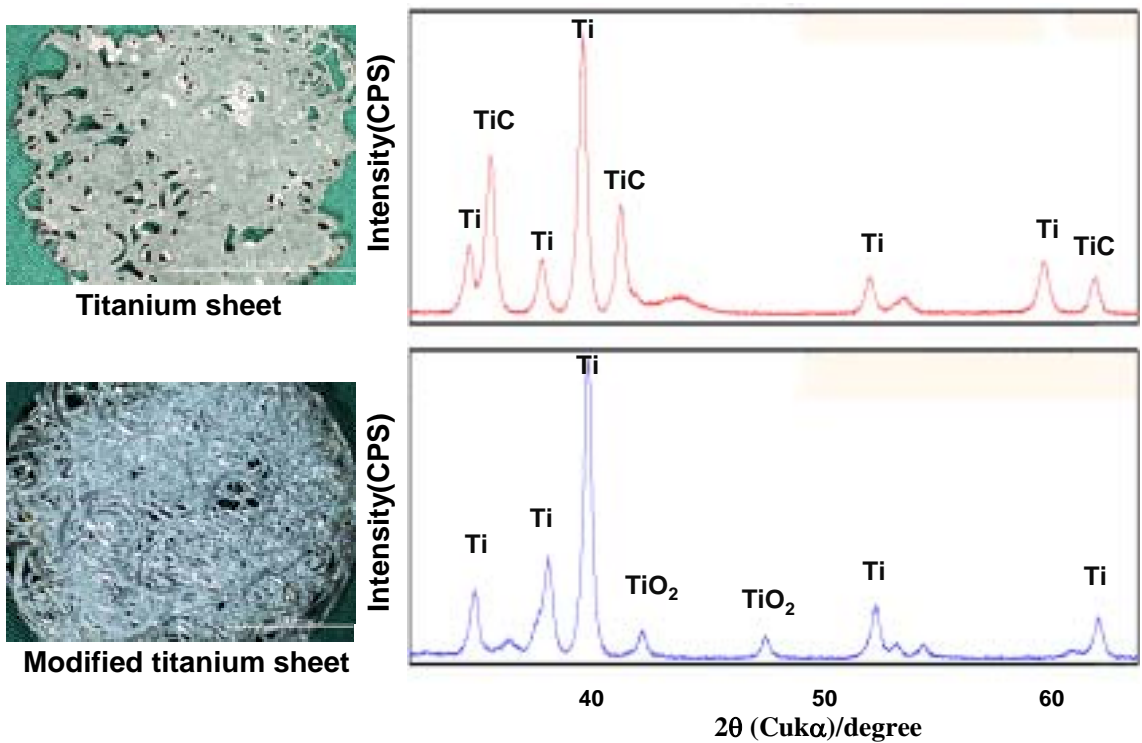

Figure 2 XRD spectra of titanium sheets processed without or with a (modified) TiN coated graphite mold.

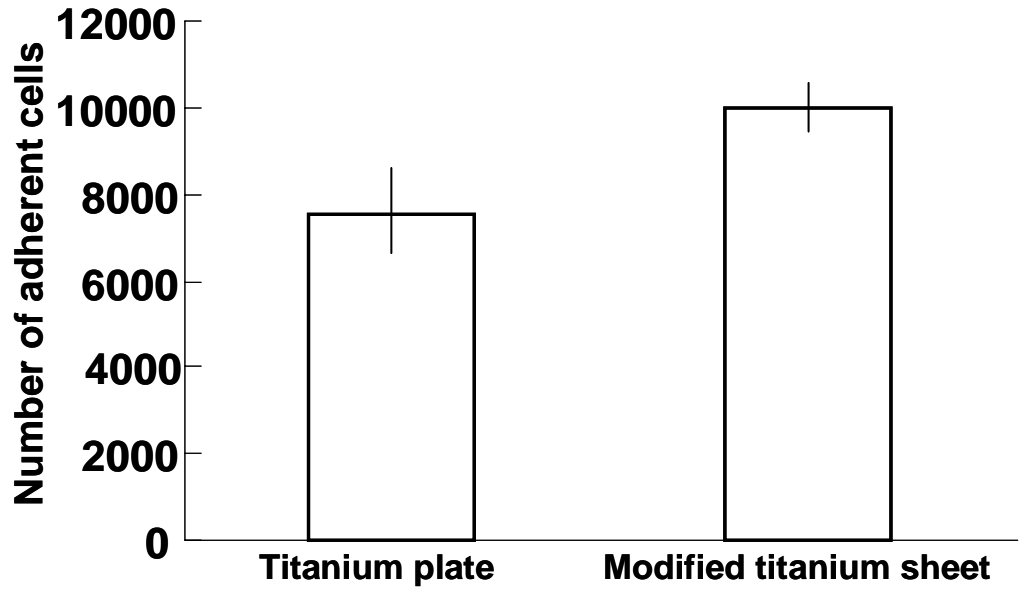

Figure 3 Number of adherent cells on titanium plate and titanium sheet processed with TiN coated graphite mold (modified) after 1d.

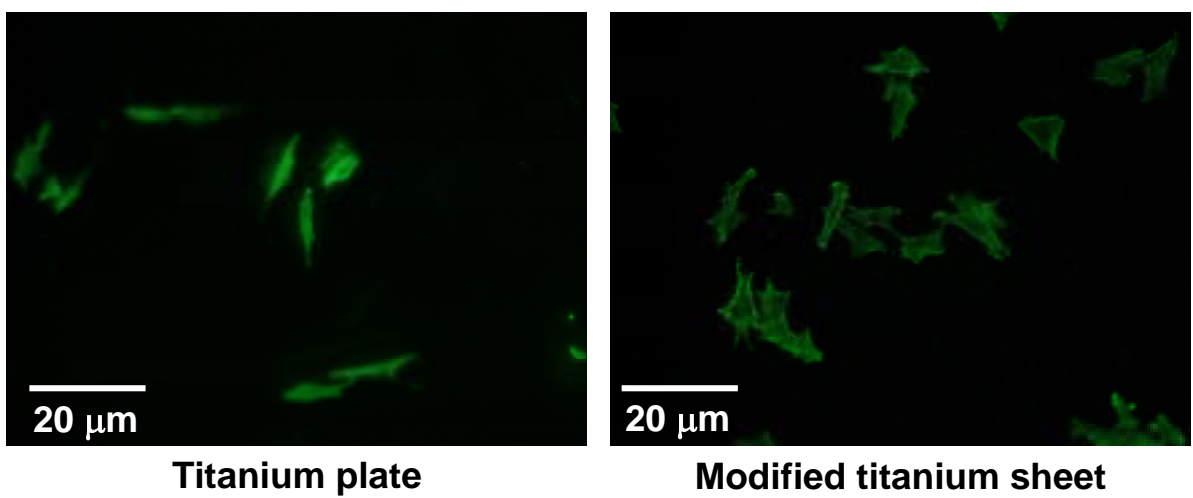

Figure 4 Fluorescence microscope images of adherent cells on titanium plate and titanium sheet processed with TiN coated graphite mold (modified).